\documentclass{eptcs}
\providecommand{\event}{ACL2-2018} 
\usepackage{underscore}           

\usepackage{tikz}
\usepackage{amsthm}
\usepackage{amssymb}
\usepackage{mathtools}
\usepackage{thmtools}
\usepackage{nameref, hyperref, cleveref}
\usepackage[title]{appendix}
\usepackage{float}
\usepackage{listings}

\usetikzlibrary{arrows.meta}

\DeclareMathOperator{\st}{st}

\crefname{equation}{}{}
\crefname{figure}{Figure}{}
\crefname{Program}{Program}{Programs}

\lstdefinestyle{progstyle}{
    aboveskip=3mm,
    belowskip=3mm,
    showstringspaces=false,
    columns=flexible,
    basicstyle={\small\ttfamily},
    numbersep=5pt,
    numberstyle=\tiny\color{gray},
    keywordstyle=\color{blue},
    commentstyle=\color{green},
    stringstyle=\color{purple},
    breaklines=true,
    breakatwhitespace=true,
    tabsize=8,
    keepspaces=true
  }

\floatstyle{ruled}
\newfloat{Program}{htb}{lop}[section]

\providecommand{\thisvolume}[1]{this volume of EPTCS, Open Publishing Association}

\theoremstyle{definition}

\newcommand{\hide}[1]{}

\title{Real Vector Spaces and the Cauchy-Schwarz Inequality in ACL2(r)}
\author{Carl Kwan \quad\quad Mark R. Greenstreet
\institute{Department of Computer Science \\
University of British Columbia\thanks{ 
 This work is supported in part by the National Science 
and Engineering
Research Council of Canada (NSERC) Discovery Grant program and
the Institute for Computing, Information and Cognitive Systems 
(ICICS) at UBC.}
\\
Vancouver, Canada}
\email{\{carlkwan,mrg\}@cs.ubc.ca}
}
\def\titlerunning{Real Vector Spaces and the Cauchy-Schwarz Inequality in ACL2(r)}
\def\authorrunning{Carl Kwan \& Mark R. Greenstreet}

\newcommand{\Rn}[1]{\ensuremath{\mathbb{R}^{#1}}}

\begin{document}
\maketitle

%
%

\begin{abstract}

We present a mechanical proof of the Cauchy-Schwarz inequality 
in ACL2(r) and a formalisation
of the necessary mathematics to undertake such a proof.
This includes the formalisation of $\mathbb R^n$ as 
an inner product space. We also provide an
application of Cauchy-Schwarz by formalising 
$\mathbb R^n$ as a metric space and exhibiting continuity for 
some simple functions $\mathbb R^n\to\mathbb R$.

The Cauchy-Schwarz inequality relates the magnitude of a vector to
its projection (or inner product) with another:
\[
|\langle u,v\rangle| \leq \|u\| \|v\|\]
with equality iff the vectors are linearly dependent. It finds frequent 
use in many branches of mathematics including linear algebra, real analysis,
 functional analysis, probability, etc. Indeed, the inequality is considered
 to be among ``The Hundred Greatest Theorems" and is listed in the
``Formalizing 100 Theorems" project. To the best of our knowledge,
our formalisation is the first published proof using ACL2(r) 
or any other first-order theorem prover.

\hide{In this paper,  we only highlight portions of interest in the 
proof and instead focus on the difficulties of such a formalisation.
For example, 
representing real vectors as lists of ACL2(r) real numbers, we needed a 
recognizer for vectors where all components are \texttt{i-small}. 
The ``obvious" recursive function is disallowed because \texttt{i-small} is
non-classical and cannot be used in the body of a recursive function. 
We provide our solutions to this and other challenges that arose in the 
course of the formalisation.}

\end{abstract}

\section{Introduction}
The Cauchy-Schwarz inequality is considered to be one of the most important inequalities in 
mathematics. Indeed, it appears in functional analysis, real analysis, probability theory, 
linear algebra,
and combinatorics to name a few. 
Cauchy-Schwarz even made an appearance on an online list of ``The Hundred Greatest Theorems"
and the subsequent formalised version of the list ``Formalizing 100 Theorems"~\cite{top100, top100formal}.
Some of the systems used for the proof include the usual suspects
HOL/Isabelle, Coq, Mizar, PVS, etc.
Notably missing, however, from the list of formalisations of Cauchy-Schwarz is a 
proof in ACL2 or ACL2(r).
We remedy this.

In this paper, we present a formal proof of the Cauchy-Schwarz inequality in
ACL2(r) including both forms (squared and norm versions) and the conditions for 
equality. 
This is the first proof of Cauchy-Schwarz 
for real vector spaces of arbitrary dimension $n\in\mathbb N$ in
ACL2(r); in fact, to the best of our knowledge, this is 
the first proof in any first-order theorem 
prover.
 Such a formalisation suggests ACL2(r) applications in the various areas 
of mathematics in which the inequality appears.
Indeed, we use Cauchy-Schwarz to prove
$\mathbb R^n$ is a metric space in this paper
and to prove 
theorems involving convex functions in \cite{Kwan2018-convex}.

The proof of Cauchy-Schwarz requires a theory of real vectors in ACL2(r). 
ACL2(r) extends ACL2 with real numbers formalised via non-standard analysis \cite{cutland},
and it supports 
automated reasoning involving irrational real and complex numbers
in addition to the respective rational subsets that are supported in the vanilla
distribution of ACL2
\cite{Gamboa2001}.
The natural next step beyond $\mathbb{R}$ is $\Rn{n}$ which is fundamental to many branches of mathematics.
Indeed, as a geometric structure, we view $\Rn{n}$ as a place in which to perform analysis;
as an algebraic object, $\Rn{n}$ is a direct sum of subspaces.
Under different lenses we view $\Rn{n}$ as different structures,
and thus it inherits the properties of the related structures.
It is a metric space, a Hilbert space, a topological space, a group, etc. 
The ubiquitous nature of $\Rn{n}$ 
suggests such a formalisation will open opportunities for applications
in the many areas of mathematics $\Rn{n}$ appears.

Formalising $\Rn{n}$ for arbitrary $n$ introduces technical difficulties in ACL2(r).
The fundamental differences between the rational and irrational numbers
induce a subtle schism between ACL2 and ACL2(r) wherein the notions
formalised in ACL2 (which are bestowed the title of \textit{classical})
are far more well-behaved than
those unique to ACL2(r) (which 
are respectively referred to as \textit{non-classical}).
The arbitrariness of $n$ suggests the necessity of defining 
 operations recursively -- yet non-classical recursive functions are not 
permitted in ACL2(r).

For the purposes of this paper, 
we present $\Rn{n}$ formalised from two perspectives. 
First, we consider $\Rn{n}$ as an inner product space
under the usual dot product.
The second perspective formalises $\Rn{n}$ as a metric space.
 In this case, we use the usual 
Euclidean metric. Metrics also provide the framework in which 
to perform analysis and we introduce notions of continuity 
in ACL2(r) for functions on $\Rn{n}$
and provide some simple examples of such functions.

Verifying the metric space properties of $\mathbb R^n$ traditionally 
requires the Cauchy-Schwarz inequality, which is the highlight of 
this paper.
Accordingly, a focus on the formalisation of Cauchy-Schwarz will 
be emphasised.
For the sake of brevity, a selection of the remaining definitions and 
theorems will be
showcased with most details omitted. 
In the remainder of the paper
we instead discuss the dynamics between approaching 
formalisation via simple definitions and avoiding fundamental logical limitations
in expressibility. For example, real vectors are unsurprisingly represented using 
lists of ACL2(r) real numbers.  However, we were surprised to find that 
deciding whether all components of a vector were infinitesimals via a
recursive function that applies a recognizer to each entry
was impossible due to the prohibition of non-standard recursive functions. We
provide our solutions to this issue and other challenges that arose in the
course of the formalisation.

Most of the mathematics encountered in this paper can be found in
standard texts such as \cite{babyrudin, shilov, lang, jacob}.
 Our proof Cauchy-Schwarz is similar to the one in \cite[Chapter 9]{roman}
and further reading specific to relevant inequalities can be found
in \cite{steele}.
Non-standard analysis is less standard of a topic but some common
references are \cite{robinson, cutland, loeb}.

\section{Related Work}
Theorem provers with 
prior formalisations of Cauchy-Schwarz include 
HOL Light, Isabelle, Coq, Mizar, Metamath, ProofPower, and PVS.
However, it appears some of the statements do not mention the 
conditions for equality~\cite{coq-cs, isabelle-cs}.

While this paper focuses on $\mathbb R^n$ from two perspectives,
most other formalisations in the literature outline only one
view of $\mathbb R^n$ -- usually either neglecting to address
metrics or lacking in the development of vectors. 
Moreover, to the best of our knowledge, these formalisations
of $\mathbb R^n$ are verified in a higher-order setting.
For, example, a  theory of complex vectors
with a nod towards applications in physics was formalised in 
HOL Light but does not address metrics \cite{Afshar}.
On the other end of the spectrum, there is a HOL formalisation
of Euclidean metric spaces and abstract metric spaces in general
that does not fully include a theory of real vectors
\cite{Harrison, Maggesi2018}. This observation extends to similar results
in Coq \cite{stein}.

Within ACL2(r), there has been formalisation for some special 
cases of $n$. The work that handled $n=1$ is indeed fundamental and
indispensable \cite{Gamboa2001}. 
We may view 
$\mathbb C\simeq \mathbb R^2$ as a vector space over $\mathbb R$ so 
$n=2$ is immediate since ACL2(r) supports complex numbers.
Moreover, extensions of $\mathbb C$ such as the quaternions 
 $\mathbb H$ and the octonions $\mathbb O$
with far richer mathematical structure than typical vector spaces
have recently been formalised, which addresses
the cases of $n=4$ and $n=8$~\cite{Cowles2017}.

\section{Preliminaries}

\subsection{Inner Product Spaces}
Inner product spaces are 
vector spaces equipped with an inner product which induces --
 among other 
notions --
rigorous definitions of the \textit{angle} between two vectors
and the \textit{size} of a vector.
More formally, a vector space is a couple $(V, F)$ where
 $V$ is a set of vectors and $F$ a field equipped with scalar 
multiplication such that
\begin{align}
&v + ( u + w) = (v + u) + w  && \textit{(associativity)} \\
&v + u = u + v && \textit{(commutativity)} \\
&\text{there exists a $0\in V$ such that $v+ 0 = v$} &&  \textit{(additive identity)} \\
&\text{there exists a $-v\in V$ such that $v + (-v) = 0$} && \textit{(additive inverse)} \\
&a(b v) = (ab)v && \textit{(compatibility)} \\
&1v = v && \textit{(scalar identity)} \\
&a(v + u) = av + au && \textit{(distributivity of vector addition)} \\
&(a + b) v = av + bv &&\textit{(distributivity of field addition)}
\end{align}
for any $v,u,w\in V$ and any $a,b\in F$
~\cite[Chapter 1]{roman}.

An inner product space is a triple $(V, F, \langle -, -\rangle)$ where
$(V,F)$ is a vector space and 
$\langle -, - \rangle : V\times V\to F$ is a function called 
an inner product, i.e. a function satisfying~
\begin{align}
&\langle a u + v,w\rangle=a\langle u,w\rangle + \langle v,w\rangle
  & \textit{(linearity in the first coordinate)}\\
&\langle u,v\rangle = \langle v, u\rangle  \text{ when } F=\mathbb R &
\textit{(commutativity)}\\
& \langle u,u\rangle \geq 0\text{ with equality iff }u = 0 & 
\textit{(positive definiteness)}
\end{align}
for any $u,v,w\in V$ and $a\in F$
\cite[Chapter 9]{roman}.

For $\Rn{n}$, this means $(\Rn{n},\mathbb R,\langle -,-\rangle)$ 
is our inner product space and we choose $\langle -,-\rangle$ to be the bilinear 
map
\begin{equation}
\langle (u_1, u_2, \hdots, u_n) , (v_1, v_2,\hdots, v_n)\rangle 
= u_1v_1 + u_2v_2+\cdots u_nv_n
\end{equation}
also known as the dot product.

\subsection{The Cauchy-Schwarz Inequality}
 Let $\|\cdot \|$ be the
norm induced by $\langle-,-\rangle$. 
 The Cauchy-Schwarz inequality states 
\begin{equation}
\tag{Cauchy-Schwarz I}
\label{ineq:cs1}
|\langle u ,v\rangle |^2 \leq \langle u,u\rangle\cdot\langle v,v\rangle
\end{equation}
or, equivalently,
\begin{equation}
\label{ineq:cs2}
\tag{Cauchy-Schwarz II}
|\langle u ,v\rangle | \leq \|u\|\|v\|
\end{equation}
for any vectors $u$, $v$ \cite[Chapter 9]{roman}.
Moreover, equality holds iff $u$, $v$ are linearly 
dependent.

\begin{proof}[Proof (sketch)]
If $v=0$, then the claims are immediate.
Suppose $v \neq 0$ and let $a$ be a field element.
Observe
\begin{equation}
\label{ineq:cs-proof}
0\leq \|u-av\|^2 = \langle u-av, u-av\rangle
=\|u\|^2- 2 a\langle u, v\rangle + a^2\|v\|^2.
\end{equation}
Setting $a=\|v\|^{-2}\langle u,v\rangle$ and rearranging produces \cref{ineq:cs1}. Take square roots and we get
\cref{ineq:cs2}.
Note that $0\leq \|u-av\|^2$ is the only step with an inequality so there is equality iff $u=av$.
\end{proof}
More details will be provided as we discuss the formal proof -- 
especially the steps that 
involve ``rearranging".

\subsection{Metric Spaces}

Metric spaces are topological spaces equipped with a metric function
which is a rigorous approach to defining the intuitive notion of 
 \textit{distance} between two vectors.
Formally,
a metric space is a couple $(M,d)$ where $M$ is a set and $d:M\times M\to \mathbb R$ a function
satisfying
\begin{align}
d(x,y) &= d(y,x) && \textit{(commutativity)} \\
d(x,y)&\geq0 \text{ with equality iff } x=y  && \textit{(positive definiteness)}\\
d(x,y) &\leq d(x,z) + d(z,y) && \textit{(triangle inequality)}
\end{align}
for any $x,y,z\in M$~\cite[Chapter 9]{roman}. The function $d$ is called a \textit{metric}.

In this case, we view $\Rn{n}$ as $(\Rn{n}, d_2)$ where
$d_2:\Rn{n} \times \Rn{n} \to \mathbb R$ denotes 
the Euclidean metric
\begin{equation}
d_2(x,y) = \sqrt {(x_1-y_1)^2+(x_2-y_2)^2+\cdots+(x_n-y_n)^2}
= \left(\sum_{i=1}^n (x_i-y_i)^2\right)^{1/2}.
\end{equation}
Proofs for the first two properties of the metric $d_2$ follow directly from the 
definition of the function.
The triangle inequality follows from the Cauchy-Schwarz 
inequality \cite[Chapter 15]{lang}.

A metric provides sufficient tools for defining continuity in a manner 
similar to that in single variable calculus \cite[Chapter 4]{babyrudin}.
A function $f:\Rn{n} \to \mathbb R$ is \textit{continuous} everywhere if 
for any $x\in\mathbb R^n$ and 
 $\epsilon>0$ there is a $\delta >0$ such that for any 
$y\in\mathbb R^n$,
if
\begin{equation}
\label{delta}
d_2(x,y) < \delta,
\end{equation}
then 
\begin{equation}
\label{eps}
|f(x) - f(y)| < \epsilon.
\end{equation}
This naturally leads to differentiability which is mentioned briefly in
\cite{Kwan2018-convex}.

\subsection{Non-standard Analysis and ACL2(r)} 

Classical real analysis is well known for its epsilon-delta approach to 
mathematical theory-building.
For example, we say that function $f: \mathbb{R} \rightarrow \mathbb{R}$ is continuous at $x \in \mathbb{R}$
iff~\cite[Chapter 4]{babyrudin}
\begin{equation}\label{eq:standard-continuity}
  \forall \epsilon > 0,\
    \exists \delta >0 : \
      \forall y\in\mathbb R, \
        | y - x | < \delta\ \implies |f(y) - f(x)| < \epsilon.
\end{equation}
%
This classical approach makes extensive use of nested quantifiers and
support for quantifiers in ACL2 and ACL2(r) is limited.  In fact, proofs involving terms with
quantifiers often involve recursive witness functions that enumerate all
possible values for the quantified term, e.g.\ see~\cite{xdoc}.
Of course, we cannot enumerate all of the real numbers.
Instead of using epsilon-delta style reasoning,
ACL2(r) is built on a formalisation of non-standard analysis
-- a more algebraic yet isomorphic approach to the theory of real 
analysis~\cite{cutland}.

Non-standard analysis introduces an extension of $\mathbb R$ called 
the \emph{hyperreals} $\ ^*\mathbb R\supset \mathbb R$ which include numbers larger in magnitude than any finite real
and the reciprocals of such numbers.
These large hyperreals are aptly named \textit{infinite} and their reciprocals are named \textit{infinitesimal}.
If $\omega$ is an infinite hyperreal, then it follows that
$\left| 1/ \omega\right| < x$
for any positive finite real $x$. Also, $0$ is an infinitesimal.

Any finite hyperreal is the sum of a real number and an infinitesimal. The 
real part of a hyperreal can be obtained through the 
\textit{standard-part} function 
$\st:\hspace{-3pt}\ ^*\mathbb R\to \mathbb R$.


To state a function $f:\Rn{n} \to \mathbb R$ is continuous in the language of non-standard
analysis amounts to:
if 
$d(x,y)$ is an infinitesimal for a standard $x$, then so is $|f(x) - f(y)|$.

In ACL2(r), lists of 
real numbers are recognized by \texttt{real-listp}. 
The recognizer for infinitesimals is the function \texttt{i-small} and 
the recognizer for infinite hyperreals is \texttt{i-large}. Reals that 
are not \texttt{i-large} are called \texttt{i-limited}.

\section{$\Rn{n}$ as an Inner Product Space in ACL2(r)}

\subsection{Vector Space Axioms}

Most of the properties of real vector spaces pass with 
relative ease and minimal guidance by the user.
Vector addition is \texttt{(vec-+ u v)} and 
scalar multiplication is \texttt{(scalar-* a v)} where $a$ is a 
field element and $u$, $v$ are vectors in $\mathbb R^n$.
The zero vector is recognized by \texttt{zvecp}.

One slightly more challenging set of theorems involve 
vector subtraction, \texttt{(vec-- u v)}.
Ideally, vector subtraction would be defined as a macro 
equivalent to \texttt{(vec-+ u (scalar-* -1 v))}
to remain consistent with
subtraction for reals in ACL2(r) and we indeed do so. 
However, it turns out that proving theorems regarding a function equivalent 
to \texttt{vec--},
which 
we call \texttt{vec--x},
and then proving the theorems for \texttt{vec--} via the equivalence 
is more amenable to verification in ACL2(r) than immediately proving theorems about 
\texttt{vec--}. In particular, the 
theorems involving closure, identity, inverses, anticommutativity, etc. 
are almost immediate using this approach.
An example of this can be seen in \cref{vec---equivalence}.
Upon verification of the desired properties for \texttt{vec--}, 
the theorem positing the
equivalence of \texttt{vec--} to \texttt{vec--x} is disabled so as to not 
pollute the space of rules.
\begin{Program}
\caption{ Using the equivalence between \texttt{vector--} and \texttt{vec---x}.
\label{vec---equivalence}}
\begin{lstlisting}
(defthm vec---equivalence
 (implies (and (real-listp vec1) 
	       (real-listp vec2)
	       (= (len vec1) (len vec2)))
	  (equal (vec-- vec1 vec2) (vec--x vec1 vec2)))
 :hints (("GOAL" :in-theory (enable vec--x vec-+ scalar-*)
		 :induct (and (nth i vec1) (nth i vec2)))))
...
(defthm vec--x-anticommutativity
	(= (vec--x vec1 vec2) (scalar-* -1 (vec--x vec2 vec1)))
 :hints (("GOAL" :in-theory (enable vec--x scalar-*))))

(defthm vec---anticommutativity
 (implies (and (real-listp vec1) (real-listp vec2) 
	       (= (len vec2) (len vec1)))
	  (= (vec-- vec1 vec2) (scalar-* -1 (vec-- vec2 vec1))))
 :hints (("GOAL" :use ((:instance vec---equivalence)
		       (:instance vec---equivalence (vec1 vec2) (vec2 vec1))
		       (:instance vec--x-anticommutativity)))))
\end{lstlisting}
\end{Program}

\subsection{Inner Product Space Axioms}

Like the vector space axioms, the majority of the relevant inner 
product space theorems 
passes by guiding ACL2(r) through textbook proofs.
Aside from the usual suspects,  one notable set of theorems are the 
bilinearity of the dot product. In particular, while the proof 
for the linearity of the first coordinate of the dot product executes via induction without  any
hints, the proof of linearity for the second coordinate does not pass so easily. 
Providing an induction scheme in the form of a hint would likely produce
the desired proof. It was simpler, however, to apply commutativity of the dot product and 
use linearity of the first coordinate to exhibit the same result, ie. given 
\begin{gather}
\langle au, v\rangle  = a\langle u,v\rangle, \\
\langle u , v \rangle = \langle v, u\rangle,
\end{gather}
we have
\begin{equation}
\langle u,av\rangle = \langle av,u\rangle = a\langle v,u\rangle = a\langle u ,v\rangle.
\end{equation}
\cref{bilinear} shows the ACL2(r) version of this proof.

\begin{Program}
\caption{An example of using commutativity to prove bilinearity of the dot product.
\label{bilinear}}
\begin{lstlisting}
(defthm dot-commutativity
 (implies (and (real-listp vec1) (real-listp vec2) 
	       (= (len vec2) (len vec1)))
	  (= (dot vec1 vec2) (dot vec2 vec1)))
 :hints (("GOAL" :in-theory (enable dot))))

(defthm dot-linear-first-coordinate-1
 (implies (and (real-listp vec1) (real-listp vec2) 
	       (= (len vec2) (len vec1)) (realp a))
	  (= (dot (scalar-* a vec1) vec2)
	     (* a (dot vec1 vec2))))
 :hints (("GOAL" :in-theory (enable dot scalar-*))))
...
(defthm dot-linear-second-coordinate-1
 (implies (and (real-listp vec1) (real-listp vec2) 
	       (= (len vec2) (len vec1)) (realp a))
	  (= (dot vec1 (scalar-* a vec2))
	     (* a (dot vec1 vec2))))
 :hints (("GOAL" :do-not-induct t
		 :use ((:instance scalar-*-closure (vec vec2))
		       (:instance dot-commutativity (vec2 (scalar-* a vec2)))))))
\end{lstlisting}
\end{Program}

Our reliance of proving theorems about algebraic structures via their algebraic properties 
instead of via induction is well exemplified here. This is especially
important for the following formalisation of metric spaces. 
Because non-classical recursive functions are not permitted in ACL2(r), 
suppressing the definition of recursive functions on vectors, say \texttt{dot},
within a \texttt{define}
facilitates the reasoning of infinitesimals in the space of $\Rn{n}$. In particular,
since $\langle -,-\rangle:\Rn{n} \to \mathbb R$ is a real-valued function, we may connect 
the notion of infinitesimal values of $\langle -, -\rangle$ with 
the entries of the vectors on which $\langle-,-\rangle $ is evaluated 
without unravelling the recursive definition of $\texttt{dot}$.
Details will be provided when we discuss the formalisation
of metric spaces.

\section{Formalising Cauchy-Schwarz}
In this section we outline some of the key lemmas in ACL2(r) that result in the Cauchy-Schwarz
inequality. Much of the proof is user guided via the algebraic properties 
of norms, the dot product, etc.
Note 
that
the lemma numbers correspond to those in the book \texttt{cauchy-schwarz.lisp}.
       By observing gaps in the numbering sequence in this presentation, the
       reader can infer where a handful of extra lemmas were needed to complete
       the proof in ACL2(r).

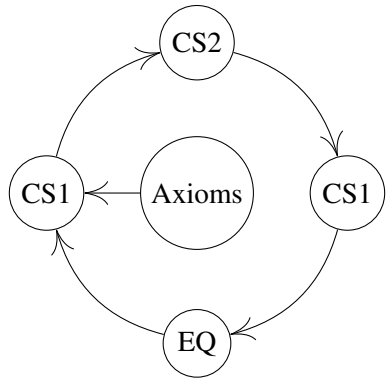
\begin{figure}[H]
\begin{center}
\caption{Structure of the proof for Cauchy-Schwarz. CS1, CS2, and EQ denotes \cref{ineq:cs1}, 
\cref{ineq:cs2}, and
the conditions for equality, respectively.}
\vspace{10pt}
\label{cs-proof-structure}
\begin{tikzpicture}[
  dot/.style={draw,circle,inner sep=3pt}]
  \node[dot] (x) at (0,2) {Axioms};
  \node[dot] (1) at (-2,2) {CS1};
  \node[dot] (2) at (0,4) {CS2};
  \node[dot] (3) at (2,2) {CS1};
  \node[dot] (4) at (0,0) {EQ};

\draw[-{>[scale = 3,length=3, width = 3]}]  (x) to (1);
\draw[-{>[scale = 3,length=3, width = 3]}]  (1) to[bend left = 30] (2);
\draw[-{>[scale = 3,length=3, width = 3]}]  (2) to[bend left = 30] (3);
\draw[-{>[scale = 3,length=3, width = 3]}]  (3) to[bend left = 30] (4);
\draw[-{>[scale = 3,length=3, width = 3]}]  (4) to[bend left = 30] (1);
\end{tikzpicture}
\end{center}
\end{figure}

\subsection{Axioms$\implies$\cref{ineq:cs1}}
Suppose $v\neq0$.
First we prove 
\begin{equation}
\label{eq:dot-id}
\| u - v\|^2 = \langle u, u \rangle - 2\langle u , v\rangle + \langle v , v\rangle
\end{equation}
by applying multiple instances of the bilinearity of $\langle - , - \rangle$. This can be seen in
\cref{prog:lem-3-4}.
Since $\|u-v\|\geq 0$, 
replacing $v$  with $\frac{\langle u,v\rangle}{\langle v ,v\rangle} v$ in 
Equation \cref{eq:dot-id} above produces
\begin{equation}
\label{ineq:used-for-reverse}
0\leq\left\| u - \frac{\langle u,v\rangle}{\langle v ,v\rangle} v\right\|^2 
= \langle u, u \rangle - 
2\left\langle u , \frac{\langle u,v\rangle}{\langle v ,v\rangle} v\right\rangle 
+ 
\left\langle \frac{\langle u,v\rangle}{\langle v ,v\rangle} v , 
	     \frac{\langle u,v\rangle}{\langle v ,v\rangle} v\right\rangle.
\end{equation}
This can be seen in \cref{prog:lem-6-7} and the inequality reduces to
\begin{equation}
 0\leq 
\langle u, u \rangle - 2\frac{\langle u,v\rangle}{\langle v ,v\rangle} \langle u , v\rangle 
+\frac{{\langle u,v\rangle}^2}{{\langle v ,v\rangle}^2} \langle v , v\rangle
=
\langle u,u\rangle 
+ \langle u,v\rangle\left(
-2\frac{\langle u,v\rangle}{\langle v,v\rangle}
+
\frac{\langle u,v\rangle}{\langle v,v\rangle}
\right)
=
\langle u,u\rangle-\frac{\langle u,v\rangle}{\langle v,v\rangle}
\end{equation}
Rearranging then produces
\cref{ineq:cs1}. 
Splitting into the 
cases $v\neq0$ and $v=0$ produces \cref{ineq:cs1} for arbitrary
vectors $u$, $v$
since the case where $v$ is zero is trivial.
This first version of 
Cauchy-Schwarz can be seen in \cref{prog:acl2-cs1}.

\begin{Program}
    \caption{Applying bilinearity of the dot product
        to prove a simple identity. \label{prog:lem-3-4}}
        \begin{lstlisting}
;; < u - v , u - v > = < u , u > - < u , v > - < v , u > + < v , v >
(defthm lemma-3
 (implies (and (real-listp u) (real-listp v) (= (len u) (len v)))
	  (equal (norm^2 (vec-- u v)) 
		 (+ (dot u u) (- (dot u v)) (- (dot v u)) (dot v v))))
 :hints (("GOAL" :use (...(:instance dot-linear-second-coordinate-2 
				     (vec1 v) (vec2 u)
				     (vec3 (scalar-* -1 v)))
		          (:instance dot-linear-second-coordinate-2 
				     (vec1 u) (vec2 u)
				     (vec3 (scalar-* -1 v))))))))

;; < u - v, u - v > = < u, u > - 2 < u , v > + < v, v >
(defthm lemma-4
 (implies (and (real-listp u) (real-listp v) (= (len u) (len v)))
	  (equal (norm^2 (vec-- u v)) 
		 (+ (dot u u) (- (* 2 (dot u v))) (dot v v))))
 :hints (("GOAL" :use ((:instance dot-commutativity (vec1 u) (vec2 v)))))))
\end{lstlisting}
\end{Program}
\begin{Program}
    \caption{Substituting $v$ for  $\langle v ,v\rangle^{-1}\langle u,v\rangle v$. \label{prog:lem-6-7}}
\begin{lstlisting}
;; 0 <= < u, u > - 2 < u , v > + < v, v >
(local (defthm lemma-6
 (implies (and (real-listp u) (real-listp v) (= (len u) (len v)))
	  (equal (<= 0 (norm^2 (vec-- u v)))
		 (<= 0 (+ (dot u u) (- (* 2 (dot u v))) (dot v v)))))))

;; let v = (scalar-* (* (/ (dot v v)) (dot u v)) v)
(local (defthm lemma-7
 (implies (and (real-listp u) (real-listp v) (= (len u) (len v))) 
	  (equal (<= 0 (norm^2 (vec-- u (scalar-* (* (/ (dot v v)) (dot u v)) v))))
	  	 (<= 0 (+ (dot u u) 
			  (- (* 2 (dot u (scalar-* (* (/ (dot v v)) (dot u v)) v))))
		 	  (dot (scalar-* (* (/ (dot v v)) (dot u v)) v)
		      	       (scalar-* (* (/ (dot v v)) (dot u v)) v))))))
 :hints (("GOAL" :use (...(:instance lemma-6 (v (scalar-* (* (/ (dot v v)) 
							     (dot u v)) v))))))))

\end{lstlisting}
\end{Program}
\begin{Program}
    \caption{Final form of \cref{ineq:cs1}. \label{prog:acl2-cs1}}
    \begin{lstlisting}
(defthm cauchy-schwarz-1
 (implies (and (real-listp u) (real-listp v) (= (len u) (len v)))
	  	 (<= (* (dot u v) (dot u v))
	      	     (* (dot u u) (dot v v))))
 :hints (("GOAL" ... :cases ((zvecp v) (not (zvecp v))))))
\end{lstlisting}
\end{Program}

\subsection{\cref{ineq:cs1}$\iff$\cref{ineq:cs2}}

To see \cref{ineq:cs2} from \cref{ineq:cs1}, we simply take square roots and 
show the equivalence between the dot products and the square of
the norms. This part can be seen in \cref{prog:acl2-cs2}.
To see the other direction, we simply square both sides and rearrange.
\begin{Program}
    \caption{Showing \cref{ineq:cs2}. \label{prog:acl2-cs2}}
    \begin{lstlisting}
(local (defthm lemma-16
 (implies (and (real-listp u) (real-listp v) (= (len u) (len v)))
	  (and (equal (acl2-sqrt (* (dot u v) (dot u v))) (abs (dot u v)))
	       (equal (acl2-sqrt (dot u u)) (eu-norm u))
	       (equal (acl2-sqrt (dot v v)) (eu-norm v))
	       (equal (acl2-sqrt (* (dot u u) (dot v v))) 
		      (* (eu-norm u) (eu-norm v)))))
 :hints (("GOAL" :use ((:instance norm-inner-product-equivalence (vec u))
		       (:instance norm-inner-product-equivalence (vec v)))))))

(defthm cauchy-schwarz-2
 (implies (and (real-listp u) (real-listp v) (= (len u) (len v)))
	  (<= (abs (dot u v))
	      (* (eu-norm u) (eu-norm v))))
 :hints (("GOAL" :use ((:instance cauchy-schwarz-1)
		       (:instance norm-inner-product-equivalence (vec v))
		       (:instance norm-inner-product-equivalence (vec u)) ...) ...)))
\end{lstlisting}
\end{Program}

\subsection{\cref{ineq:cs1}$\iff$Conditions for Equality}

Suppose $u,v\neq 0$ and
\begin{equation}
\langle u, v\rangle ^2 = \langle u, u\rangle \langle v, v\rangle.
\end{equation}
Then we simply reverse all the equalities used to prove
\cref{ineq:cs1} from the axioms (see Inequality \cref{ineq:used-for-reverse})  until we return to
\begin{equation}
0 = \left\|u -\frac{\langle u,v\rangle}{\langle v ,v\rangle} v \right\|^2.
\end{equation}
Since $\|\cdot\|^2$ is positive definite, we must
have 
\begin{equation}
u =\frac{\langle u,v\rangle}{\langle v ,v\rangle} v.
\end{equation}
This can be seen in \cref{prog:linearity}.

To show linearity, we introduce a Skolem function so that the final form
of the conditions for equality are in the greatest generality. 
We also attempted a proof where the value of the Skolem constant
        was explicitly computed -- 
simply find the first non-zero element
    of $v$ and divide the corresponding element of 
$u$ by the $v$ element.
    The proof for the Skolem function approach was much simpler
    because the witness value comes already endowed with
     the properties we need for subsequent
    reasoning. In particular, invoking linear dependence to show 
\cref{ineq:cs1} simply amounts to applying algebraic 
rules to an arbitrary unknown witness which is simple 
(though occasionally tedious) from our 
formalisation. Otherwise, not 
using a Skolem function would necessitate the exhibition of
a particular coefficient to complete the implication graph
in \cref{cs-proof-structure}.
The definition
of the Skolem function is in \cref{prog:skolem}. The cases 
for $u=0$ or $v=0$ are immediate.
The final result can be seen in \cref{prog:cs1-equality}.
The conditions for equality for \cref{ineq:cs2} follows from its equivalence to 
\cref{ineq:cs1}.

\begin{Program}
    \caption{Linearity follows from equality.\label{prog:linearity}}
    \begin{lstlisting}
(local (defthm lemma-19
 (implies (and (real-listp u) (real-listp v) (= (len u) (len v)) (not (zvecp v))
	       (= (* (dot u v) (dot u v)) (* (dot u u) (dot v v))))
	  (equal u (scalar-* (* (/ (dot v v)) (dot u v)) v))) ...))
\end{lstlisting}
\end{Program}
\begin{Program}
    \caption{Introducing a Skolem function for linearity.\label{prog:skolem}}
    \begin{lstlisting}
(defun-sk linear-dependence-nz (u v)
 (exists a (equal u (scalar-* a v))))
\end{lstlisting}
\end{Program}
\begin{Program}
\caption{The  conditions
    for equality for \cref{ineq:cs1}.\label{prog:cs1-equality}}
\begin{lstlisting}
(defthm cauchy-schwarz-3
 (implies (and (real-listp u) (real-listp v) (= (len u) (len v)))
	  (equal (= (* (dot u v) (dot u v)) (* (dot u u) (dot v v)))
		 (or (zvecp u) (zvecp v) (linear-dependence-nz u v)))) ...)
\end{lstlisting}
\end{Program}

\subsection{Final Statement of the Cauchy-Schwarz Inequality}
\cref{prog:final} displays our final form of Cauchy-Schwarz.
The ACL2(r) theorems \texttt{cauchy-schwarz-1} and
\texttt{cauchy-schwarz-2}
are equivalent to \cref{ineq:cs1} and \cref{ineq:cs2}, respectively.
The theorems \texttt{cauchy-schwarz-3} and \texttt{cauchy-schwarz-4}
correspond to the conditions for equality for 
\cref{ineq:cs1} and \cref{ineq:cs2}, respectively.
\begin{Program}
    \caption{The final form of Cauchy-Schwarz.\label{prog:final}}
    \begin{lstlisting}
(defthm cauchy-schwarz-1
 (implies (and (real-listp u) (real-listp v) (= (len u) (len v)))
	  	 (<= (* (dot u v) (dot u v))
	      	     (* (dot u u) (dot v v)))) ...)

(defthm cauchy-schwarz-2
 (implies (and (real-listp u) (real-listp v) (= (len u) (len v)))
	  (<= (abs (dot u v))
	      (* (eu-norm u) (eu-norm v)))) ...)

(defthm cauchy-schwarz-3
 (implies (and (real-listp u) (real-listp v) (= (len u) (len v)))
	  (equal (= (* (dot u v) (dot u v))
	      	    (* (dot u u) (dot v v)))
		 (or (zvecp u) (zvecp v) 
		     (linear-dependence-nz u v)))) ...)

(defthm cauchy-schwarz-4
 (implies (and (real-listp u) (real-listp v) (= (len u) (len v)))
	  (equal (=  (abs (dot u v)) (* (eu-norm u) (eu-norm v)))
	  	 (or (zvecp u) (zvecp v) 
		     (linear-dependence-nz u v)))) ...)
\end{lstlisting}
\end{Program}

\section{$\Rn{n}$ as a Metric Space in ACL2(r)}

\subsection{Metric Space Axioms}
Observe
\begin{equation}
d_2(x,y) = \| x - y\|_2 = \sqrt {\langle x - y, x - y\rangle}.
\end{equation}
Proving theorems regarding metrics reduces to proving theorems
about the norm from which the metric is induced.
Likewise, proving theorems involving norms can be reduced to 
proving properties about the underlying inner product.
The process of formalisation, then, should ideally define 
the metric via the norm, and the norm should be defined via the inner product.
This is useful for proving properties such as positive definiteness of 
both the metric and the norm, eg. 
if 
\begin{equation}
\langle u, u \rangle \geq 0
\end{equation} 
with equality iff $u=0$, then the 
same applies for 
\begin{equation}
\| x\|_2 = \sqrt{\langle x, x\rangle}\geq0
\end{equation}
and so
\begin{equation}
d_2(x,y)=\| x-y\|_2 = \sqrt{\langle x-y, x-y\rangle}\geq0.
\end{equation}

However, as exemplified by
\texttt{vec--} and \texttt{vec--x}, the obvious sequence is not necessarily the easiest.
Indeed, not only is it simpler to prove theorems on functions equivalent to 
the desired functions, we also prove properties of similar functions not equivalent to the 
desired functions but such that if the properties hold for the similar functions, then
 they also hold for the desired functions.
For example, suppose we wish to prove commutativity for the Euclidean metric 
\texttt{eu-metric}. 
Recall \texttt{(vec-- x y)} is a macro for \texttt{(vec-+ x (scalar-* -1 y))}
and, together with an instance of \texttt{acl2-sqrt}, formalising 
the proofs for commutativity 
require a non-trivial amount of hints and user guidance.
We instead define recursively a function \texttt{metric\textasciicircum2},
which is the square of the norm of the difference of two vectors 
(which is equivalent to the square of the Euclidean metric,
 i.e.\ $\|x-y\|_2^2$).
Moreover, proving those equivalences simply amounts to unwinding the definitions.
Having established
\begin{equation}
\|x-y\|^2_2 = \|y - x\|^2_2,
\end{equation}
it follows that 
\begin{equation}
d_2(x,y)= \sqrt{\|x-y\|^2_2 }= \sqrt{\|y - x\|^2_2} = d_2(y,x).
\end{equation}
Hence, commutativity for \texttt{eu-metric} is proven.
\begin{Program}
\caption{The Euclidean norm, metric, and squared metric. \label{eu-metric}}
\begin{lstlisting}
(defun eu-norm (u)
 (acl2-sqrt (dot u u)))
...
(defun eu-metric (u v)
 (eu-norm (vec-- u v)))
...
(define metric^2 (vec1 vec2) ...
 (norm^2 (vec-- vec1 vec2)) ...)
\end{lstlisting}
\end{Program}


\subsection{Continuity $\Rn{n}\to \mathbb R$}
To showcase continuity, let us begin with an enlightening example.
Recall the non-standard analysis definition of 
continuity for a function $f:\Rn{n}\to \mathbb R$ 
stated in the language of non-standard analysis:
if $d_2(x,y)$ is an infinitesimal for a standard $x$, then so is 
$f(x) - f(y)$. 
Take $f(x)= \sum_{i=1}^n x_i$.
It is clear that $f$ is continuous in our usual theory of classical real analysis. 
However, we must translate this into the language of infinitesimals.
By hypothesis,
\begin{equation}
\label{hyp}
d_2 (x,y) = \|x-y\|_2 = \sqrt{(x_1-y_1)^2 + (x_2-y_2)^2 + \cdots  + (x_n-y_n)^2}
\end{equation}
is an infinitesimal. We would like to show that 
\begin{equation}
\label{wts}
f(x) - f(y) = \sum_{i=1}^n x_i - \sum _{i=1}^n y_i = \sum_{i=1}^n (x_i - y_i)
\end{equation}
is also an infinitesimal. 
Indeed, by Equation~\cref{hyp} we see that each $x_i-y_i$ must 
necessarily be 
infinitesimal since otherwise $d_2(x,y)$ wouldn't be an infinitesimal.
Because the RHS of Equation~\cref{wts} is a finite sum of infinitesimals, 
so is $f(x)-f(y)$ as desired. 
The two motivating questions are: 
\begin{enumerate}
\item How do we make ACL2(r) recognize $x_i-y_i$ are infinitesimals from $d_2(x,y)$ being
infinitesimal?
\item How do we state ``all $x_i-y_i$ are infinitesimals"?
\end{enumerate}

To answer the first question, observe for any vector $z$ and  $i\leq n$,
\begin{equation}
\|z\|_2=\sqrt{\sum_ {i=1}^n z_i^2} \geq \max _i |z_i|\geq |z_i|.
\end{equation}
Setting $z=x-y$, we see that if the norm is an infinitesimal, then so must each entry of 
the vector.
By introducing an ACL2(r) function, say \texttt{max-abs-reals}, equivalent to $\max_i$
and reasoning over the arbitrariness of $i$ instead of over the length of $x$ and $y$, 
we may exhibit the infinitesimality of any entry in $x-y$ as seen in Program \ref{max}.

\begin{Program}
\caption{Showing arbitrary entries of a vector are infinitesimal via the maximum element.
\label{max}}
\begin{lstlisting}
(define max-abs-reals ((vec real-listp))
 ...
 (b* (((unless (consp vec)) 0)
       ((cons hd tl) vec)
       ((unless (realp hd)) 0))
      (max (abs hd) (max-abs-reals tl)))...)

(defthm eu-norm-i-small-implies-max-abs-reals-i-small
 (implies (and (real-listp vec) (i-small (eu-norm vec)))
          (i-small (max-abs-reals vec))))

(defthm eu-norm-i-small-implies-elements-i-small
 (implies (and (real-listp vec) (i-small (eu-norm vec)) (natp i) (< i (len vec)))
          (i-small (nth i vec))))
\end{lstlisting}
\end{Program}

To address the second question, one could imagine a recognizer for vectors 
with infinitesimal entries -- such a recognizer is depicted in \cref{i-small-vecp}.
However, this recognizer would be recursive on the entries of the vector with each recursive
step invoking the non-classical recognizer \texttt{i-small} for infinitesimal reals.
Because non-classical recursive functions are forbidden, so is the suggested recognizer.
A Skolem function was also considered as a possibility but
to remain consistent with \texttt{eu-norm-i-small-implies-elements-i-small}
a theorem positing the condition for an arbitrary index $i$ as seen in
\cref{i-small metric} was chosen instead.
\begin{Program}
\caption{A fantastical recognizer for vectors with infinitesimal entries that doesn't exist.
\label{i-small-vecp}}
\begin{lstlisting}
(defun i-small-vecp (x)
 (cond ((null x) t)
       ((not (real-listp x)) nil)
       (t (and (i-small (car x)) (i-small-vecp (cdr x))))))
\end{lstlisting}
\end{Program}
\begin{Program}
\caption{A theorem positing the infinitesimality of arbitrary entries in $x-y$. 
\label{i-small metric}}
\begin{lstlisting}
(defthm eu-metric-i-small-implies-difference-of-entries-i-small
 (implies (and (real-listp x) (real-listp y) (= (len y) (len x)) 
	       (natp i) (< i (len x)))
               (i-small (eu-metric x y)) 
          (i-small (- (nth i x) (nth i y)))) ...)
\end{lstlisting}
\end{Program}

To see \texttt{eu-metric-i-small-implies-difference-of-entries-i-small} in action,
consider once again the example of $f(x) = \sum_{i=1}^n x_i$.
If $n=3$ and \texttt{sum} is the ACL2(r) function equivalent to $f$, then the following is 
the proof of continuity for \texttt{sum}.
Other examples of functions with proofs of continuity in ACL2(r) include 
the Euclidean norm, the dot product with one coordinate fixed, 
and the function $g(x,y) = xy$.
\begin{Program}
\caption{A theorem positing \texttt{sum} is continuous.\label{sum-is-continuous}}
\begin{lstlisting}
(defthm sum-is-continuous
 (implies (and (real-listp x) (real-listp y) (= (len x) 3) (= (len y) (len x)) 
	       (i-small (eu-metric x y)))
          (i-small (- (sum x) (sum y))))...)
\end{lstlisting}
\end{Program}

\section{Conclusion}

Firstly, we would like to note the choice of classical proof on which we base this
formalisation. In particular, we would like to compare its flavour to other 
proofs of Cauchy-Schwarz. Indeed, there are geometric proofs, analytical proofs,
combinatorial proofs, inductive proofs, etc.~\cite{wu2009} 
whereas we followed a rather algebraic approach. Considering ACL2(r)'s strengths 
with regards to induction, the choice may seem odd. Indeed, there are 
several potential inductive candidates we considered at the onset of this endeavor
before proceeding with the algebraic approach.
However, most of the other candidates inducted over the dimension of $\mathbb R^n$ 
and required reasoning over the real entries of vectors.
We suspect unwinding the vectors and guiding ACL2(r) through 
such a proof would be more onerous than the one outlined in this paper. 
Moreover, our formalisation of inner 
product spaces already provided the exact tools necessary 
for our preferred proof  of Cauchy-Schwarz (eg. vectors, 
vector-vector operations, scalar-vector operations,
inner products, etc.)
without resorting to reasoning over individual reals.
The precision of this approach, while arguably more elegant,
also complements our approach to defining continuity. 
By reasoning over the properties of the vector itself instead 
of the individual components, we circumvent the 
soundness-motivated limitations of ACL2(r).

Secondly, 
this formalisation  has two notable purposes.
The first is the various applications that may be introduced as 
a result of the Cauchy-Schwarz inequality. 
The appearance of Cauchy-Schwarz in 
  functional analysis, real analysis, probability theory, 
combinatorics, etc. speaks to its utility.  A further application of 
Cauchy-Schwarz is in \cite{Kwan2018-convex} where a set of theorems
involving convex functions are formalised.
Additionally, 
the various structures of $\Rn{n}$ are very rich in mathematical  theory
and hold applications in various areas of science. 
In this paper, we presented a formalisation of the space from
only two perspectives. However, the choice of perspectives is arguably among
the most fundamental. It is the vector space structure of $\Rn{n}$ that
provides the necessary operations between its elements. Indeed, one would be 
hard-pressed to find any view of $\Rn{n}$ that does not assume operations on 
the domain. Moreover, inner products are the path to calculus: 
inner products lead to norms; norms lead to metrics; metrics 
lead to real analysis. 
The formalisation of $\Rn{n}$ as a metric space is the last step before 
multivariate calculus, which is in of itself highly applicable and left
as future work. 
We also note the possibility of proving Cauchy-Schwarz for more
abstract structures as future work.

During the course of formalisation, emphasis was placed on using algebraic methods 
to prove theorems that would have otherwise been proved via induction. However, 
algebraic approaches require significant guidance from the user. 
Instead, the properties of the inner product space axioms 
and the proof of Cauchy-Schwarz might be amenable 
to certification by a SMT solver via 
\texttt{smtlink}~\cite{Peng2015,Peng2015-NASA}.
The challenge here is that SMT solvers do not perform
induction -- we need to leave that for ACL2.
On the other hand, we might be able to treat operations on vectors as 
uninterpreted functions
with constraints corresponding to the requirements for a function 
to be an inner product, a norm, etc.

Finally, we discuss further formalisations of $\Rn{n}$ under different lenses.
In fact, there is still potential to further extend $\Rn{n}$ as a metric space.
 The notions of continuity are independent of the metric used and
$d_2$ may be replaced with any metric on $\Rn{n}$. By way of 
encapsulation, pseudo-higher-order techniques may be employed to easily 
formalise various real metric spaces -- especially if we consider the 
metric induced by other $p$-norms. 

Among the extensions of $\Rn{n}$ as a metric space is proving its completeness.
Addressing Cauchy sequences traditionally follows from an application of 
Bolzano-Weierstrass which has yet to be formalised in 
ACL2~\cite{babyrudin}.
Stating completeness in
terms infinitesimals and ACL2(r) is a farther
but tantalizing prospect. Upon doing so, we would have a formalisation of $\Rn{n}$ 
as a Hilbert space~\cite{riesz}.

\bibliography{refs}
\bibliographystyle{eptcs}

\begin{appendices}
\section{Why are Non-classical Recursive Functions Prohibited?}
A discussion with Ruben Gamboa\footnote{
We would like to thank Ruben Gamboa for his insightful explanations 
on which most of this appendix is based.}
 and members of the ACL2 Help Mailing List 
sheds light on why non-classical recursive functions are prohibited. 
In summary, the introduction of such functions will also introduce 
inconsistency into the logic of ACL2(r).
It is possible to define a function that violates the rules
of non-standard analysis. 
\begin{Program}
\caption{An impossible function in ACL2(r).\label{contra}}
\begin{lstlisting}{language=Lisp}
(defun f (n)
 (cond ((zp n) 0)
       ((standardp n) n)
       (t (f (-1 n)))))
\end{lstlisting}
\end{Program}

For example, 
consider the hypothetical function defined in \cref{contra}.
Suppose \texttt f terminates.
If \texttt n is standard, then \texttt f returns it without issue.
If \texttt n is infinite, then \texttt f returns the largest standard number. 
However, this is impossible since such a number does not exist 
and, if \texttt n is infinite, so is \texttt {(-1 n)} which means \texttt f 
should not have terminated anyways. 
Moreover, this applies if \texttt n is any non-standard hyperreal since 
if $n= \st(n) + \epsilon$ is equivalent to \texttt n and
$\epsilon>0$ is an infinitesimal,
then $n-1 = \st(n-1) + \epsilon$ is not standard either. 

The essence of the issue with \texttt f is that its measure is non-standard. If the measure of 
a function can be proven to be standard, then a recursive non-classical function could 
be conceded. However, in practice, such a proof would likely be subtle and involved.
\end{appendices}

\end{document}